\documentclass[%
reprint,
superscriptaddress,
 amsmath,amssymb,
 aps,
]{revtex4-2}
\usepackage{graphicx}% Include figure files
\usepackage{dcolumn}% Align table columns on decimal point
\usepackage{bm}% bold math
\usepackage{hyperref}% add hypertext capabilities
\usepackage[mathlines]{lineno}% Enable numbering of text and display math
\usepackage{natbib}
\usepackage{jabbrv}
\begin{document}
\citestyle{nature}
%\preprint{APS/123-QED}

%\title{Multiphoton Quantum Interference in Ultracompact Inverse-Designed Multiport Beam Splitter}%
\title{Multiphoton Quantum Interference at Ultracompact Inverse-Designed \mbox{Multiport Beam Splitter}}%

\author{Shiang-Yu Huang}
\email{shiang-yu.huang@fmq.uni-stuttgart.de}
\author{Shreya Kumar}
\author{Jeldrik Huster}
%\affiliation{\mbox{Institute for Functional Matter and Quantum Technologies, University of Stuttgart, 70569 Stuttgart, Germany}} %Lines break automatically or can be forced with \\
\affiliation{Institute for Functional Matter and Quantum Technologies, University of Stuttgart, 70569 Stuttgart, Germany} %Lines break automatically or can be forced with \\

\author{Yannick Augenstein}
\affiliation{Institute of Theoretical Solid State Physics, Karlsruhe Institute of Technology, 76131 Karlsruhe, Germany}

\author{Carsten Rockstuhl}
\email{carsten.rockstuhl@kit.edu}
\affiliation{Institute of Theoretical Solid State Physics, Karlsruhe Institute of Technology, 76131 Karlsruhe, Germany}
\affiliation{Institute of Nanotechnology, Karlsruhe Institute of Technology, 76131 Karlsruhe, Germany}

\author{Stefanie Barz}
\email{stefanie.barz@fmq.uni-stuttgart.de}
% \affiliation{\mbox{Institute for Functional Matter and Quantum Technologies, University of Stuttgart, 70569 Stuttgart, Germany}}
% \affiliation{\mbox{Center for Integrated Quantum Science and Technology (IQST), University of Stuttgart, 70569 Stuttgart, Germany}}
\affiliation{Institute for Functional Matter and Quantum Technologies, University of Stuttgart, 70569 Stuttgart, Germany}
\affiliation{Center for Integrated Quantum Science and Technology (IQST), University of Stuttgart, 70569 Stuttgart, Germany}



\begin{abstract}
Photonic quantum technologies enter a new phase when realized in photonic integrated circuits, leading to a great advance in practical applications.
In the pursuit of high integration density and low circuit complexity, ultracompact devices delivered by topology optimization offer a promising solution to miniaturize these photonic systems even further.
However, their potential for quantum experiments has not yet been fully explored despite the constant development.
In this work, we demonstrate multiphoton quantum interference using a topology-optimized tritter with a size of 8.0~\textmu m~\texttimes~4.5~\textmu m. 
We characterize the tritter and reconstruct its transfer matrix by means of single- and two-photon statistics.
We also perform heralded three-photon quantum interference with the tritter. 
The measured four-fold coincidence features a peak with visibility of ($-47.9\pm 8.6$)\%, which is in fair agreement with the prediction of $-55.8$\% estimated from the reconstructed transfer matrix.
Our work confirms successful multiphoton quantum interference at an ultracompact interferometer and demonstrates the possibility of utilizing topology-optimized multiport interferometers in various fields of quantum technologies.
\end{abstract}

\maketitle

%\tableofcontents

\section*{Introduction}
Quantum technologies exploit the non-classical properties of quantum systems and hold great promise to enhance communication security~\cite{scarani2009security,lo2014secure} and computational speed of specific tasks, 
such as boson sampling~\cite{tillmann2013experimental,spring2013boson}, prime factorization~\cite{shor1994algorithms,lu2007demonstration,politi2009shor} and simulation of quantum systems~\cite{georgescu2014quantum,daley2022practical}.
Among a variety of physical implementations, photons are a competitive candidate since they are resilient to quantum decoherence due to their weak interaction with the environment.
Such a property is advantageous to the realization of quantum technologies as the photonic quantum states can be generated and manipulated under normal ambient conditions.
Despite the fact that single photons do not directly interact with each other in the scheme of linear optics \cite{kok2000limitations,kok2007linear}, 
their interference together with projective measurements offers a way to entangle them, generating the desired photonic quantum states via a probabilistic process \cite{pan2012multiphoton}.

As a fundamental component enabling photon interference in an optical setup, interferometers play a central role in manipulating the photonic quantum states.
The most basic interferometer in an optical system is usually a 2~\texttimes~2 beam splitter.
In combination with phase shifters, such beam splitters can be arranged in specific configurations, known as the Reck \cite{reck1994experimental} and Clements schemes \cite{clements2016optimal}, to implement arbitrary linear operations for manipulating photonic quantum states. 
In the case of bulk optics, multiphoton entangled quantum states, such as the Greenberger–Horne–Zeilinger (GHZ) state~\cite{lu2014experimental,wang2016experimental, zhong201812, cao2024photonic} and the W state \cite{eibl2004experimental, tashima2009local},
can be produced via either heralding or post-selection of the single photons.
\begin{figure}[t!]
  \centering
  \includegraphics[scale=0.31]{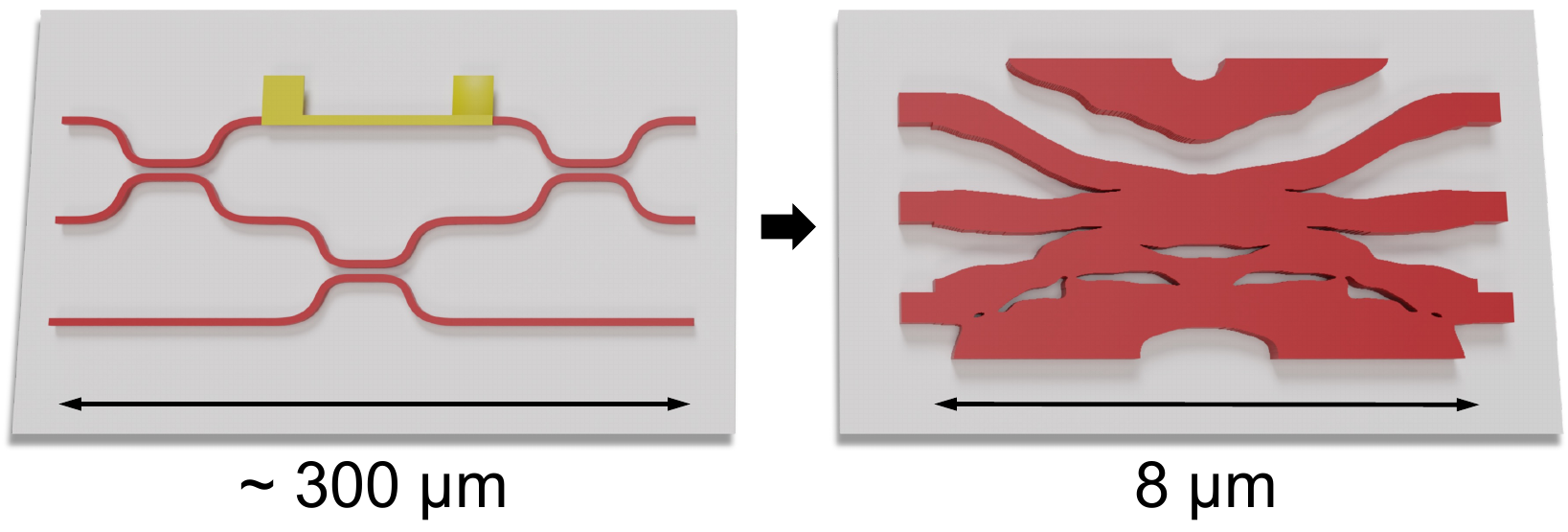}
  \caption{
    Topology optimization reduces the spatial footprint of the tritter device by at least a factor of 30. 
    Left: conventional tritter composed of 2~\texttimes~2 beam splitters and a phase shift. Right: topology-optimized tritter (not to scale).
    }
    \vspace{-1em}
  \label{fig:tritter circuit comparison}
  \end{figure}
Furthermore, beam splitters can be expanded to multiport ones, namely the input and output modes $>$~2, for sophisticated applications and thus gaining advantages.
For instance, applying multiport interferometers for constructing photonic circuits is in theory more resource-efficient and resilient to system errors~\cite{arends2024decomposing}.
In addition, multiport interferometers give rise to unique photon statistics or nontrivial features of multiphoton quantum interference~\cite{bhatti2023generating, bhatti2024heralding}.
Pioneering investigations have mainly focused on quantum experiments using a fiber-based tritter, i.e., a 3 \texttimes~3 beam splitter constructed by 3 equidistant optical fibers to achieve equal field coupling strength.
With such a tritter, researchers have demonstrated multiphoton interference~\cite{spagnolo2013three, menssen2017distinguishability, spring2017chip,kim2021implementation} and the generation of the W state, G state and GHZ state~\cite{kumar2023experimental}.

In the pursuit of large-scale quantum applications, photonic integrated circuits emerge as a promising platform due to their scalability as well as their thermal and mechanical stability.
Essential operations required by quantum technologies can be implemented with a minimal footprint.
For instance, with integrated interferometers realized by directional couplers or multimode interferometers (MMI), researchers have demonstrated the on-chip generation of Bell states \cite{lee2024quantum}, GHZ state \cite{pont2024high,lee2024quantum}, and multiphoton graph states \cite{adcock2019programmable, vigliar2021error}.
There has also been growing interest in direct implementation of integrated multiport interferometers in photonic integrated systems.
Such an integrated device has been experimentally demonstrated as a 4~\texttimes~4~MMI, which enables non-classical interference between two single photons~\cite{peruzzo2011multimode}.
Still, due to the self-imaging effect that relates to the operating wavelength~\cite{capmany2020programmable}, MMIs usually occupy a large footprint (ranging from $\sim$100~\textmu m to $\sim$1,800~\textmu m in length) and are not ideal for realizing a compact multiport interferometer.
In this context, inverse design via topology optimization is advantageous as it provides a route to design integrated components with minimal spatial footprint.  
Elements of photonic integrated systems, such as fiber-to-chip grating couplers~\cite{dory2019inverse, hammond2022multi, pita2022ultracompact, wang2024single}, wavelength-dependent demultiplexers~\cite{vercruysse2019analytical, piggott2020inverse}, and polarization rotator splitter \cite{chen2024integrating},  
have been redesigned using topology optimization, achieving comparable or superior performance yet with a much smaller spatial footprint compared to the conventional counterparts.
A previous publication has also demonstrated two-photon interference with topology-optimized 2~\texttimes~2~beam splitters~\cite{he2023super}.
However, quantum interference of more than two photons within an ultracompact multiport interferometer still remains unexplored.

In this work, we perform multiphoton quantum experiments with a topology-optimized tritter fabricated on the silicon-on-insulator (SOI) platform.
The dimensions of the on-chip tritter are 8.0~\textmu m \texttimes~4.5~\textmu m, which is a great improvement in the spatial footprint compared to other types of tirtters (Fig.~\ref{fig:tritter circuit comparison}).
We perform two-photon and heralded three-photon interference using spontaneous parametric down-conversion (SPDC) single-photon sources and reconstruct the transfer matrix of the tritter to predict the non-classical photon statistics.
The difference between the complete set of the measured two-photon Hong-Ou-Mandel (HOM) visibilities and that of the predicted ones is $5.54\%$.
The visibility of the four-fold coincidence measurement for three-photon interference yields a value of ($-47.9\pm8.6$)\%, which is in fair agreement with the predicted one of $-55.8$\% calculated from the reconstructed transfer matrix.
Overall, our work shows a promising potential of topology-optimized multiport interferometers for photonic quantum applications and offers an element that could construct an integrated photonic architecture with rich functionality but lower layout complexity.
\section*{Results}
\begin{figure*}[t!]
  \centering
  \includegraphics[scale=0.37]{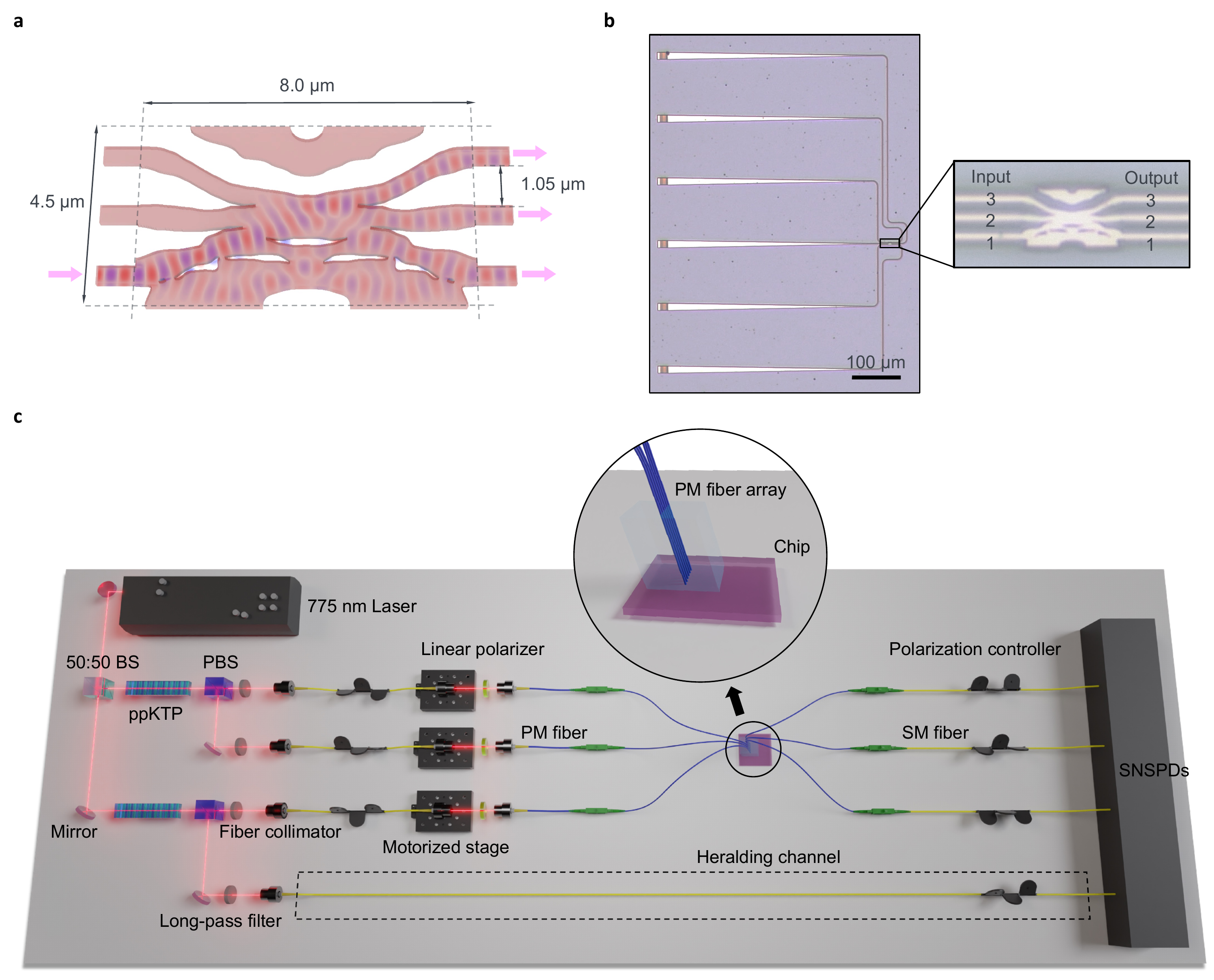}
  \caption{
    The topology-optimized tritter and the experimental setup for the multiphoton interference.
    (a) Sketch of the designed topology-optimized tritter on the SOI platform (the cladding layers and the substrate are omitted). 
    The dimensions of the interference region are 8.0~\textmu m~\texttimes~4.5~\textmu m. 
    The spacing between the input and output waveguides is 1.05 \textmu m. 
    The dimensions of the silicon input and output waveguides are 450~nm~\texttimes~220~nm.
    (b) Optical microscope images of the fabricated integrated photonic circuit with the topology-optimized tritter.
    (c) Schematic of the experimental setup for the quantum interference with the topology-optimized tritter (not to scale).
    BS: Beam splitter.
    PBS: Polarizing beam splitter.
    PM fiber: Polarization-maintaining fiber.
    SM fiber: Single-mode fiber.
    SNSPD: Superconducting nanowire single photon detector.
    Note that one, two or three single photons can be sent to the chip by blocking or swapping the corresponding input fiber channels.
    }
    \vspace{-1em}

  \label{fig:tritter structure and exp setup}
  \end{figure*}

\begin{figure*}[t!]
  \centering
  \includegraphics[scale=0.38]{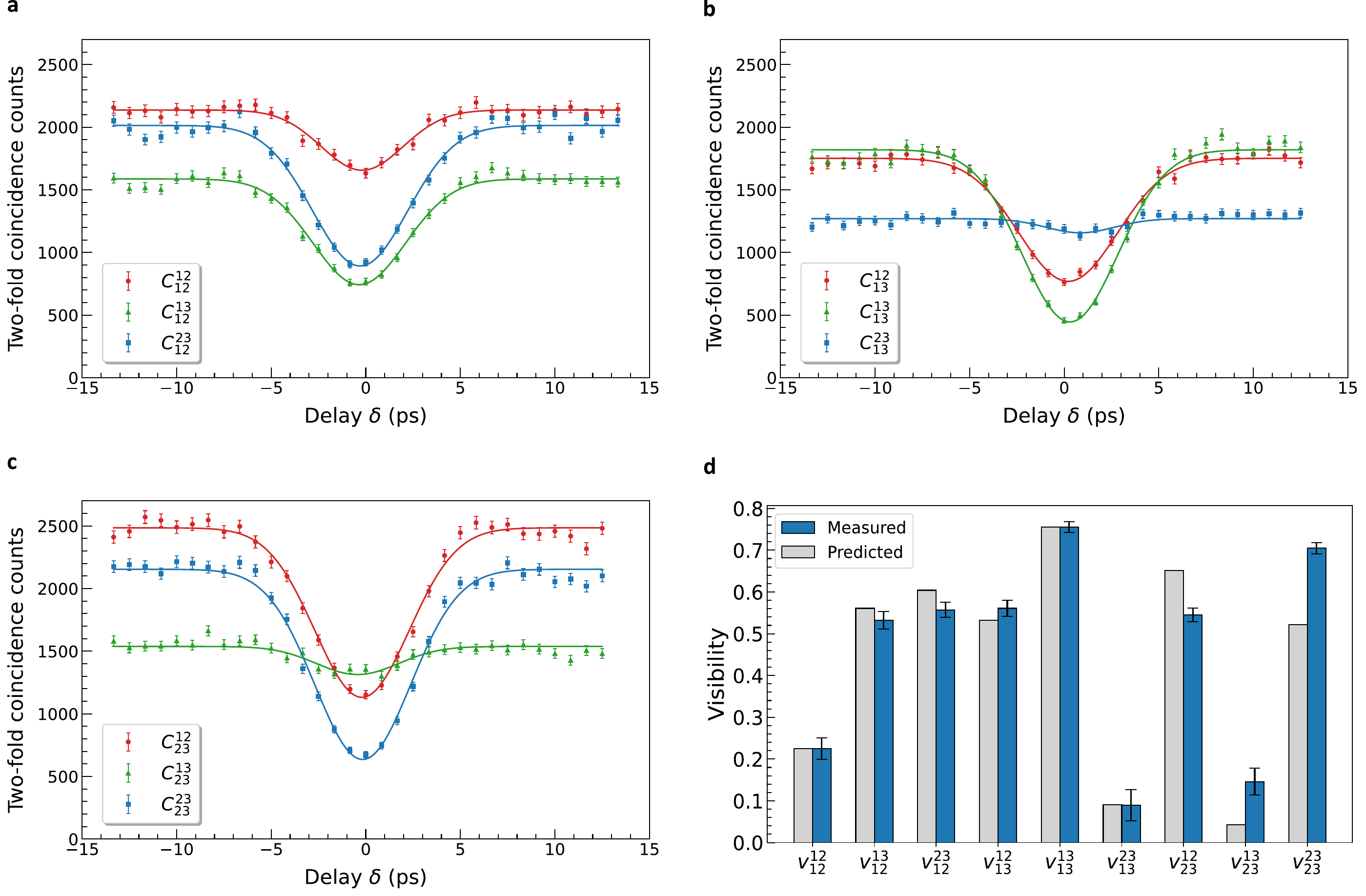}
  \caption{
    Results of the two-photon experiments.
    (a) The input modes of the tritter $i, j = 1,2$. (b) $i, j = 1,3$. (c) $i, j = 2,3$.
    Here, $C_{ij}^{ml}$ in the legends denotes the two-photon coincidence counts with the input modes $i, j$ and output modes $l, m$. 
    Data points and the corresponding Gaussian fitting are represented by dots with error bars and lines, respectively.
    The integration time of each data point is 60 seconds.
    (d) Measured two-photon HOM visibilities $V_{ij}^{lm}$ and their theoretical prediction from the reconstructed transfer matrix, where $i, j$ ($k, l$) indicate the input (output) modes of the tritter.
    }
  \label{fig:2-photon data}
  \end{figure*}

\subsection*{Device design and simulations}
To leverage the mature complementary metal-oxide-semiconductor (CMOS) manufacturing technologies and operate at the telecom wavelength, we inversely design a tritter on the 220-nm SOI platform via topology optimization.
The entire design process is carried out using the open-source software package Meep \cite{oskooi2010meep, hammond2022high}, which solves the Maxwell equations using the finite-difference time-domain (FDTD) method.
In need of an ultracompact photonic integrated interferometer, the design region, where the optical wave or photon interference takes place, is confined within an area of 8.0~\textmu m \texttimes~4.5 \textmu m.
The single-mode strip waveguides connected to the input and output ports are spaced surface to surface by 1.05 \textmu m.

The goal of the inverse design is to optimize the device geometry such that its transfer matrix matches a target one, which in this case is that of an ideal tritter \cite{zukowski1997realizable}
\begin{equation}
  %\scriptsize
  \normalsize
  \mathbf{M}_{\rm{tritter}} = 
  \frac{1}{\sqrt{3}}
  \begin{pmatrix}
    1 & 1  & 1 \\
    1 & e^{i\frac{2\pi}{3}} & e^{-i\frac{2\pi}{3}} \\
    1 & e^{-i\frac{2\pi}{3}}  & e^{i\frac{2\pi}{3}}
  \end{pmatrix}.
  \label{eq: ideal tritter transfer matrix}
\end{equation}
For each input mode $i$, the figure of merit (FoM) is defined by 
\begin{equation}
  {\mathrm{FoM}}_i = \Bigl\lvert\sum_{l} \bigl(M^*_{li} \alpha_{l}\bigr)\Bigr\rvert,
  \label{eq: FoM}
\end{equation}
where $l$ indicates the output mode, $M_{li}$ the corresponding target matrix element, and $\alpha_{l}$ the complex amplitude of the photonic mode in the output waveguide $l$ after the excitation travels through the input mode $i$.
Here, the complex overlap between $M_{li}$ and $\alpha_{l}$ takes into account the amplitude and phase of the matrix elements, which ensures that the optimization shapes the device's transfer matrix toward $\mathbf{M}_{\mathrm{tritter}}$ rather than merely maximizing transmission efficiency.
Considering the number of the tritter input modes, in total three separate FoM functions are applied in the optimization process (additional details in Materials and methods).

Eventually, the topology optimization delivers a reflectively symmetric structure that features hollows in the bottom area and slender paths interfacing the input and output ports (Fig.~\ref{fig:tritter structure and exp setup}(a)).
The optical responses of the final design are further investigated by performing FDTD simulations using Ansys Lumerical FDTD.
The simulation results suggest that the tritter works as expected, and only the splitting ratio of the topology-optimized tritter is slightly uneven (Fig.~S1(b)-(d) in Supplementary Information).
Its insertion loss varies from approximately $7.8\%$ to $12.7\%$ at the wavelength of 1550 nm, depending on which input port the light source is located.
The insertion loss is seemingly attributed to the strong optical resonance in the gaps between the slender paths (Fig.~S2(a)-(c) in the Supplementary Information). 
Light could be confined in these cavity-like regions and then scattered upward or downward, leading to additional loss of the device.
After the investigation, the topology-optimized tritter is fabricated via standard foundry fabrication process.
The three input and output ports are connected by six grating couplers for interfacing optical fibers (Fig.~\ref{fig:tritter structure and exp setup}(b)). 
With this configuration, the photonic integrated circuit corresponds to what has been considered in the design.

\subsection*{Characterization and multiphoton experiments}
We perform on-chip quantum interference utilizing single photons generated by SPDC sources.
In the experiments, a pulsed laser at the wavelength of 775 nm pumps the periodically poled titanyl phosphate (ppKTP) crystal, generating pairs of single photons at the wavelength of 1550 nm  (Fig.~\ref{fig:tritter structure and exp setup}(c)).
The photon pairs are separated by polarizing beam splitters (PBSs) and then coupled into single-mode (SM) fibers with free-space fiber collimators. 
The pump laser is blocked using long-pass filters.
The polarization of the photons is manipulated by polarization controllers and the residual polarization is cleaned using linear polarizers for coupling the photons into the polarization-maintaining (PM) fibers of the fiber array.
The distinguishability of the single photons is introduced by the motorized stages that move the free-space fiber collimators.
Relative time delays of the photons can, therefore, be swept by consecutively shifting the motorized stages.
For the on-chip experiments, single photons are coupled into and out from the topology-optimized tritter via a single PM fiber array and fiber-to-chip grating couplers.
After the chip, the photons are filtered by band-pass filters with a bandwidth of 1.5 nm and detected by superconducting nanowire single-photon detectors (SNSPDs) with a detection efficiency of around 90\%.

As it is critical to reveal the linear operation the topology-optimized tritter performs on the injected single photons, 
we reconstruct its transfer matrix and analyze the relevant multiphoton statistics accordingly.
Specifically, this is achieved via the measurements using single photons and the quantum interference between two photons \cite{laing2012super}.
To begin with, single photons from one SPDC source are injected into each of the tritter input modes $i \in \{1,2,3\}$ individually and the photon counts at the output modes $l \in \{1,2,3\}$ are recorded for retrieving the amplitude of the elements in the reconstructed transfer matrix.
Next, paired single photons from one SPDC source are sent to two of the three tritter input modes in all possible pairwise combinations for performing two-photon HOM interference.
The relative time delay $\delta$ is swept by moving the corresponding motorized states and the integration time of each data point is 60 seconds.
The resulting two-fold coincidence counts exhibit the characteristic dips (Fig.~\ref{fig:2-photon data}(a-c)).
According to the following definition
\begin{equation}
  V_{ij}^{lm} = \frac{C_{ij}^{lm}(\delta=\infty) - C_{ij}^{lm}(\delta=0)}{C_{ij}^{lm}(\delta=\infty)},
  \label{eq:visibility}
\end{equation}
the value of the HOM visibilities is estimated from the Gaussian fittings of the data points (blue bars in Fig.~\ref{fig:2-photon data}(d)).
Here, $C_{ij}^{lm}$ indicates the two-fold coincidence counts and $i, j$ ($l, m$) denote the input (output) modes of the tritter. 
Compared to the visibilities predicted using Eq.~\ref{eq: ideal tritter transfer matrix}, in which $V_{ij}^{lm} = 0.5$ for all port combinations, it is obvious that the topology-optimized tritter does not function as a balanced one.
Based on the data sets acquired from the single-photon detection and the two-photon HOM interference, the algorithm stated in Ref.~\citenum{laing2012super} produces the transfer matrix of the topology-optimized tritter 
\begin{equation}
  \scriptsize
  %\normalsize
  \mathbf{M}_{\rm{rec.}} = 
  \frac{1}{\sqrt{3}}
  \begin{pmatrix}
    1.016 e^{-i0.036\pi} & 0.995 e^{i0.059\pi}  & 1.039 e^{-i0.018\pi} \\
    1.013 e^{-i0.058\pi} & 0.904 e^{i0.610\pi} & 0.686 e^{-i0.748\pi} \\
    1.022 e^{i0.020\pi} & 0.699 e^{-i0.577\pi}  & 1.199 e^{i0.814\pi}
  \end{pmatrix}, 
  \label{eq: tritter transfer matrix}
\end{equation}
with an average 1\% and 4.5\% deviation in the estimated amplitude and the phase terms of the elements, respectively (additional details shown in S2 in Supplementary Information).
From Eq.~\ref{eq: tritter transfer matrix}, we can predict the two-photon visibilities $V_{ij}^{lm}$ by computing the probabilities of the corresponding output states of the two-photon interference (gray bars in Fig.~\ref{fig:2-photon data}(d)).
In the comparison, all measured values generally follow the predictions despite some slight mismatch in the magnitude of the visibilities (Fig.~\ref{fig:2-photon data}(d)), which is likely due to the variation in the acquired data sets.
To quantify the effectiveness of the reconstructed transfer matrix, we calculate the difference between the measured and predicted HOM visibilities \cite{tillmann2016unitary}
\begin{align}
  Q_{\rm{vis.}} = \frac{1}{n} \sum^{n}_{} |{V^{\rm{meas.}} - V^{\rm{pred.}}} |,
  \label{eq: difference of visibilities}
  \end{align}
where $n=9$ is the total number of the HOM visibilities, $V^{\rm{meas.}}$ the measured HOM visibilities and $V^{\rm{pred.}}$ the predicted ones.
The calculation using Eq.~\ref{eq: difference of visibilities} gives rise to $Q_{\rm{vis.}} = 5.54\%$, suggesting an appropriate reconstruction of the tritter transfer matrix.

Single photons give rise to diverse detection patterns when traveling through a multiport interferometer, which reflects the device's capability for essential quantum operations.
Exploring the unique three-photon statistics of the topology-optimized tritter is, therefore, crucial and the analysis represents the first step towards its practical use in a quantum photonic circuit.
To perform the three-photon quantum experiment, three photons are injected into all input modes of the tritter, respectively, and one additional photon is heralded (Fig.~\ref{fig:tritter structure and exp setup}(c)).
After the photons pass through the chip, the four-fold coincidence counts of all detection channels are measured.
The distinguishability of the photons is introduced by moving the motorized stage that connects to input 1 of the tritter, varying the relative time delay $\delta$.

\begin{figure}[t]
  \centering
  \includegraphics[scale=0.38]{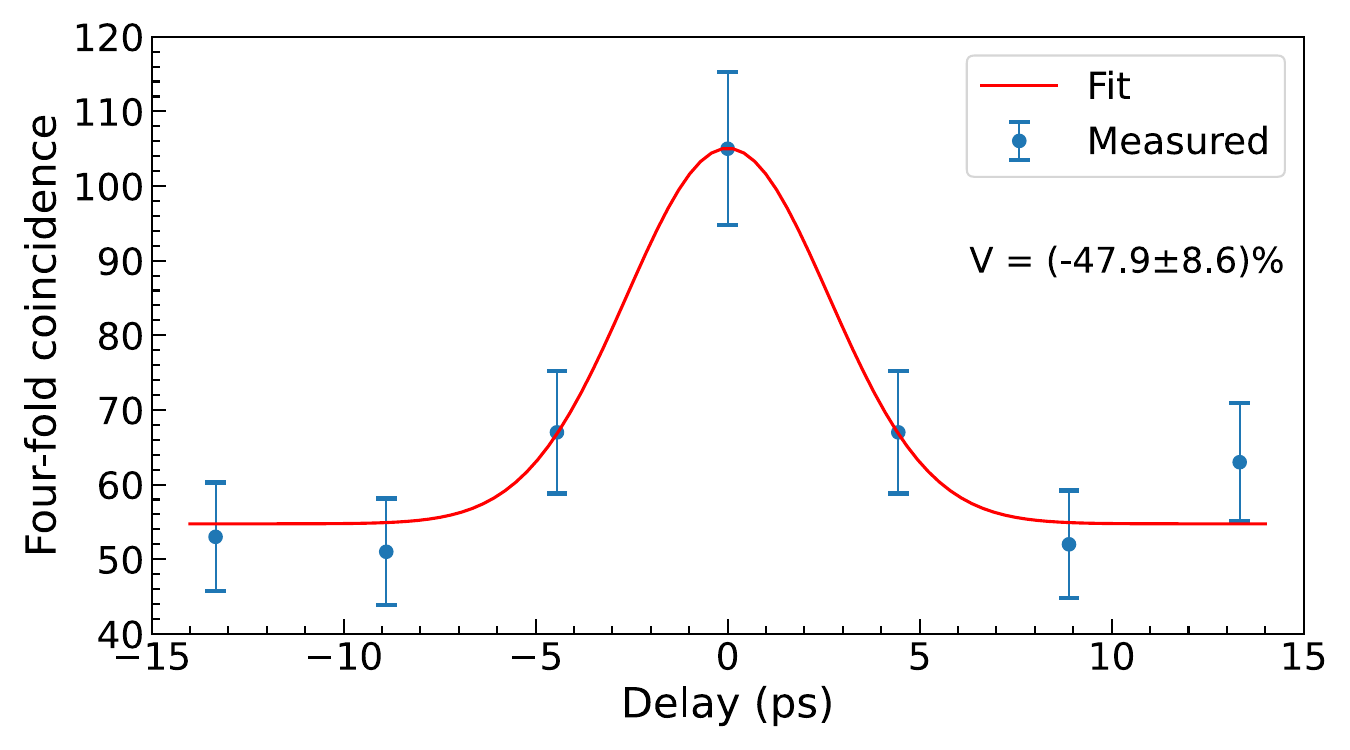}
  \caption{
  Four-fold coincidence measurement with the topology-optimized tritter.
    The measured visibility of the peak is estimated to be $(-47.9\pm8.6)$\%, which is in fair agreement with the predicted value of $-55.8$\%.
    The overall integration time of the measurement is roughly 30 days for 7 data points with different time delays.        
    }
  \label{fig:4-fold data}
  \end{figure}

The visibility of the three-photon interference is defined by 
\begin{equation}
  V_{ijk}^{lmn} = \frac{C_{ijk}^{lmn}(\delta=\infty) - C_{ijk}^{lmn}(\delta=0)}{C_{ijk}^{lmn}(\delta=0)},
  \label{eq: three-fold visibility}
\end{equation}
where $i,j,k$ ($l,m,n$) denote the input (output) modes of the tritter.
Based on Eq.~\ref{eq: three-fold visibility}, the predicted visibility $V_{123}^{123}$ calculated from the reconstructed transfer matrix in Eq.~\ref{eq: tritter transfer matrix} is $-55.8\%$,
which is in fair agreement with the measured one of $(-47.9\pm 8.6)\%$ estimated from the Gaussian fit of the experimental data (Fig.~\ref{fig:4-fold data}).
Note that, according to Eq.~\ref{eq: three-fold visibility}, negative visibility implies a peaked curve in the coincidence counts, which is observed in the experiment results.
The data rate of the four-fold coincidence is suboptimal due to the limited coupling efficiency of the chip.
The total integration time of the measurement is around 30 days for 7 data points.

\section*{Discussion}
In summary, we demonstrate multiphoton quantum interference, namely the two-photon HOM effect and three-photon interference, with an ultracompact on-chip tritter designed using topology optimization.
The spatial footprint of the tritter is significantly reduced compared to other types of multiport integrated interferometers.
The transfer matrix that describes the tritter's linear operation is reconstructed from the measurement data and is used to predict multiphoton statistics.
The three-photon interference is investigated by measuring the four-fold coincidence of all detection channels and the resulting visibility aligns well with the prediction.

This successful demonstration of multiphoton interference at the topology-optimized tritter highlights its applicability and potential for various quantum technologies.
For instance, the topology-optimized tritter could be used to create qutrits for efficient quantum computation by reducing the number of gate operations \cite{lanyon2009simplifying} or for other high-dimensional quantum applications, such as the measurement of three-dimensional Bell states or high-dimensional quantum teleportation \cite{luo2019quantum,hu2020experimental}, in an on-chip fashion.
In addition, utilizing this type of ultracompact multiport interferometers in photonic integrated circuits may improve the integration density and lower the complexity of large-scale photonic integrated circuits, 
potentially leading to reduced insertion loss, lower resource requirement and higher resilience to noise~\cite{arends2024decomposing}.
Overall, we anticipate that the development of photonic quantum technologies benefits from the utilization of the topology-optimized tritter.

\section*{Materials and methods}
\label{sec:Materials and methods}
\subsection*{Topology optimization}
The optimization task is addressed by the adjoint-based algorithm that calculates the gradient of the FoM functions per port with respect to the design density $\rho$, which represents the permittivity in the discretized design region. 
Then, we sum over the gradient contributions from all ports and renew $\rho$. 
This leads to an update of the permittivity distribution for maximizing the FoM functions towards one.
With a suitable projection that forces an increasing binarization of $\rho$ in the course of the optimization, 
the permittivity of each pixel in the design region ends up taking only two different values, reflecting the presence of two selected materials, Si and SiO$_2$.
In the filtering process performed in each iteration of the optimization, we convolve $\rho$ with a Gaussian kernel to remove high-spatial-frequency or checkerboarding features, which implicitly reflects the minimal spatial resolution achievable with our fabrication technique. 
This process delivers spatial distributions of the material in the design region that ensures the desired functionality.
The initial condition within the design region is set to be a slab of a pseudo material whose permittivity is the average of Si and SiO$_2$.
Eventually, the optimization undergoes 407 iterations with a duration of around 131 hours and the overall FoM reaches 96.9\%.

\subsection*{Device fabrication}
The layout of the photonic circuit with the topology-optimized tritter is sent to Institut f\"ur Mikroelektronik Stuttgart (IMS Chips) for standard fabrication. 
Specifically, the tritter structure is fabricated on an SOI wafer with a 220-nm device layer and a 3-\textmu m bottom oxide layer.
The photoresists are first patterned using the electron-beam lithography method and then the tritter structure is transferred to the device layer through one full-etch process.
A 1-\textmu m top oxide layer is later deposited using the low pressure chemical vapor deposition (LPCVD) process.

\subsection*{Acknowledgments}
  S.-Y.H., S.K., J.H. and S.B. would like to acknowledge the support from the Carl Zeiss Foundation, the Centre for Integrated Quantum Science and Technology (IQST), the Federal Ministry of Education and Research (BMBF, projects SiSiQ: FKZ 13N14920, PhotonQ: FKZ 13N15758, QRN: FKZ16KIS2207), and the Deutsche Forschungsgemeinschaft (DFG, German Research Foundation, 431314977/GRK2642).
  J.A. and C.R. would like to acknowledge the support from the Deutsche Forschungsgemeinschaft (DFG, German Research Foundation) under Germany’s Excellence Strategy via the Excellence Cluster 3D Matter Made to Order (EXC-2082/1-390761711) and the Carl Zeiss Foundation via the CZF-Focus@HEiKA Program.
  The authors would also like to thank Simone d'Aurelio for the fruitful discussion and thank Jonas Zatsch and Louis Hohmann for the technical support for the experiments.
%\end{acknowledgments}

%\bibliographystyle{jabbrv_abbrv} 
\bibliography{achemso-demo}% Produces the bibliography via BibTeX.

\end{document}

%\preprint{APS/123-QED}

\title{Supplementary Information: Multiphoton Quantum Interference at Ultracompact Inverse-Designed Multiport Beam Splitter}%

\author{Shiang-Yu Huang}
\email{shiang-yu.huang@fmq.uni-stuttgart.de}
\author{Shreya Kumar}
\author{Jeldrik Huster}
\affiliation{Institute for Functional Matter and Quantum Technologies, University of Stuttgart, 70569 Stuttgart, Germany} %Lines break automatically or can be forced with \\

\author{Yannick Augenstein}
\affiliation{Institute of Theoretical Solid State Physics, Karlsruhe Institute of Technology, 76131 Karlsruhe, Germany}

\author{Carsten Rockstuhl}
\email{carsten.rockstuhl@kit.edu}
\affiliation{Institute of Theoretical Solid State Physics, Karlsruhe Institute of Technology, 76131 Karlsruhe, Germany}
\affiliation{Institute of Nanotechnology, Karlsruhe Institute of Technology, 76131 Karlsruhe, Germany}

\author{Stefanie Barz}
\email{stefanie.barz@fmq.uni-stuttgart.de}
\affiliation{Institute for Functional Matter and Quantum Technologies, University of Stuttgart, 70569 Stuttgart, Germany}
\affiliation{Center for Integrated Quantum Science and Technology (IQST), University of Stuttgart, 70569 Stuttgart, Germany}



\maketitle

%\tableofcontents
\newpage
\section{Simulated Optical Responses of the Topology-Optimized Tritter}
\begin{figure*}[h!]
  \centering
  \includegraphics[scale=0.37]{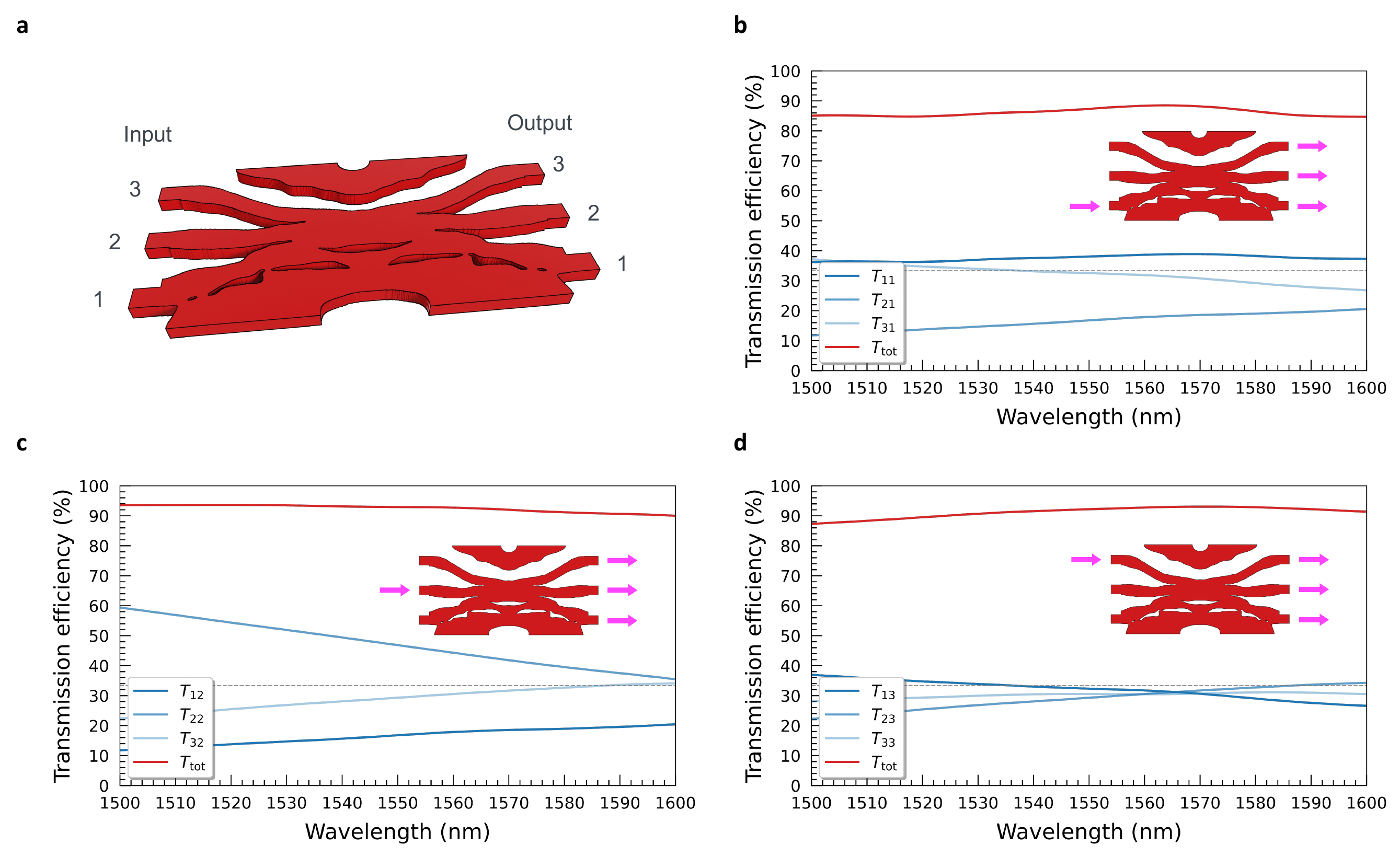}
  \caption{Simulated transmission efficiency of the topology-optimized tritter.
  (a) Sketch of the tritter input and output modes.
  The transmission efficiency, which is determined by the overlap integral of the input mode and the fundamental TE mode of the waveguide, is simulated when the light source is at (b) input 1, (c) input 2, and (d) input 3.
  $T_{li}$ in the legends indicates the corresponding transmission efficiency, where $i$ ($l$) denotes the input (output) mode of the tritter.
  $T_{\rm{tot}}$ is the sum of the three corresponding $T_{li}$.
  The grey dashed line indicates the 1/3 splitting ratio.
  }
  \label{fig:transmission efficiency}
  \end{figure*}

  \begin{figure}[h]
    \centering
    \includegraphics[scale=0.37]{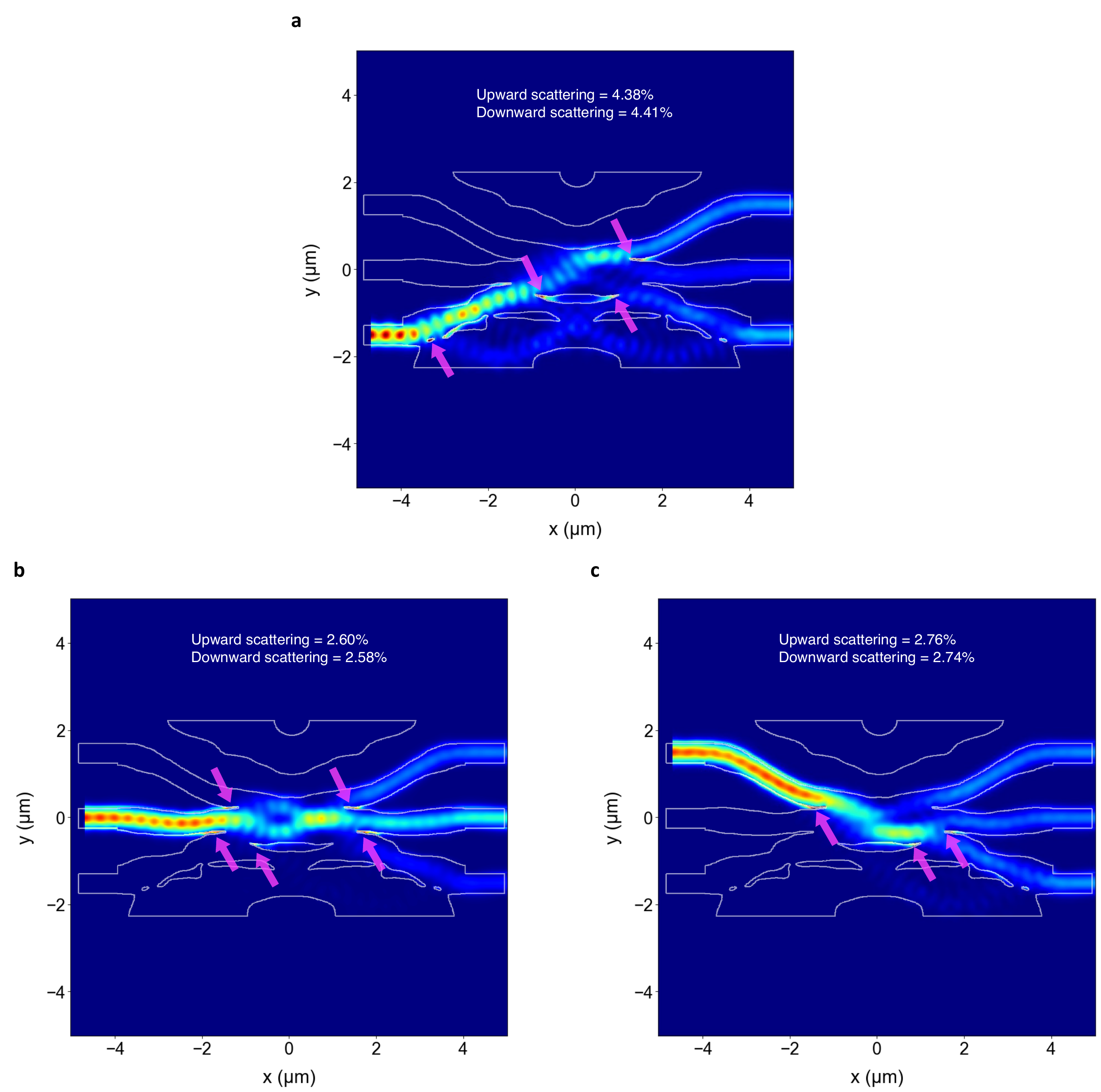}
    \caption{Simulated intensity distribution at 1550 nm in the topology-optimized tritter when the light source is at (a) input 1, (b) input 2, and (c) input 3.
    The optical resonance occurs in the small gaps between the paths and the edge of the hollows, leading to "hot spots" in the small areas (magenta arrows).
    The amount of the corresponding upward and downward light scattering are indicated in the figures.
    }
    \label{fig:intensity distribution}
    \end{figure}

\newpage
\section{Analysis of the reconstructed tritter transfer matrix}
\begin{equation}
  \scriptsize
  %\normalsize
  M_{\rm{rec.}} = 
  \frac{1}{\sqrt{3}}
  \begin{pmatrix}
    (1.016\pm0.004) e^{-i(0.036\pm0.002)\pi} & (0.995\pm0.004) e^{i(0.059\pm0.003)\pi}  & (1.039\pm0.004) e^{-i(0.018\pm0.002)\pi} \\
    (1.013\pm0.004) e^{-i(0.058\pm0.003)\pi} & (0.904\pm0.006) e^{i(0.610\pm0.008)\pi} & (0.686\pm0.006) e^{-i(0.748\pm0.008)\pi} \\
    (1.022\pm0.004) e^{i(0.020\pm0.002)\pi} & (0.699\pm0.004) e^{-i(0.577\pm0.010)\pi}  & (1.199\pm0.009) e^{i(0.814\pm0.006)\pi}
  \end{pmatrix}.
  \label{eq: tritter transfer matrix with errors}
\end{equation}
The transfer matrix of the topology-optimized tritter is reconstructed using the method presented in Ref.~[39] in the main text, which requires the measured single-photon counts and the two-photon visibilities.
The deviation of the reconstructed matrix elements (Eq.~\ref{eq: tritter transfer matrix with errors}) are then estimated using the Monte Carlo method, where Poissonian noise is introduced to the measured datasets and the tritter transfer matrix is reconstructed 200 times.
By calculating the standard deviation of the elements in the 200 reconstructed matrices, we acquire on average 1\% and 4.5\% deviation in the estimated amplitude and the phase terms of the matrix elements.

%%%------
%\newpage

%\bibliography{achemso-demo}% Produces the bibliography via BibTeX.